\documentclass[pra,twoside,amsfonts,amssymb,amsmath,titlepage,floatfix,letterpaper,twocolumn]{revtex4}
\usepackage{amsmath}
\usepackage{amsfonts}
\usepackage{amssymb}
\usepackage{graphicx}
\usepackage{graphicx,array}
\usepackage{dcolumn}

\newcommand{\rr}[1]{\mathrm{#1}}

\newcommand{\bellst}[4]{\ensuremath{(\hat{a}_{ #1 }^{\dag} \hat{b}_{ #2 }^{\dag}
 \, - \, \hat{a}_{ #3 }^{\dag} \hat{b}_{ #4 }^{\dag})}}

 \newcommand{\bellstpb}[4]{\ensuremath{(\hat{a}_{ #1 }^{\dag} \hat{b}_{ #2 }^{\dag}
 \, + \, e^{i \theta}\hat{a}_{ #3 }^{\dag} \hat{b}_{ #4 }^{\dag})}}
  \newcommand{\bellstpa}[4]{\ensuremath{( e^{i \theta} \hat{a}_{ #1 }^{\dag} \hat{b}_{ #2 }^{\dag}
 \, + \, \hat{a}_{ #3 }^{\dag} \hat{b}_{ #4 }^{\dag})}}

\begin{document}

\title{Optimal Generation of Pulsed Entangled Photon Pairs}

\author{Juan F. Hodelin}
\email[corresponding author: ]{hodelin@physics.ucsb.edu}

\author{George Khoury}

\author{Dirk Bouwmeester}

\affiliation{Department of Physics, University of California,
Santa Barbara, CA 93106, USA}

\date{\today}

\begin{abstract}
We experimentally investigate a double-pass parametric
down-conversion scheme for producing pulsed,
polarization-entangled photon pairs with high visibility. The
amplitudes for creating photon pairs on each pass interfere to
compensate for distinguishing characteristics that normally
degrade two-photon visibility. The result is a high-flux source of
polarization-entangled photon pulses that does not require
spectral filtering. We observe quantum interference visibility of
over 95\% without the use of spectral filters for 200 femtosecond
pulses, and up to 98.1\% with 5 nm bandwidth filters.
\end{abstract}

\maketitle

\section{Introduction}
%
Quantum entanglement is one of the most striking concepts in
quantum physics, and has now become the primary resource in the
fields of quantum communication and quantum computation. Quantum
information systems utilize entanglement to transmit, encode, and
process data stored on qubits. For quantum communication protocols
the natural candidate for a qubit is the photon.  Qubits encoded
in photons can fly freely through space or travel over optical
fibers for long distances without decoherence.

Time of arrival~\cite{engtime,timebin}, spatial mode
~\cite{rarity}, and polarization~\cite{ou1,shih} are some of the
degrees of freedom in which photons have been entangled.
Polarization-entangled photon pairs have been produced using
parametric down-conversion (PDC) reliably for almost two
decades~\cite{ou1,shih}. A convenient source of
polarization-entangled photons without the need for post-selection
is non-collinear Type-II PDC~\cite{kwiatpdc}. These
polarization-entangled photons have been used for many experiments
including violations of Bell's inequality~\cite{kwiatpdc}, quantum
teleportation~\cite{teleportation}, quantum
cloning~\cite{cloning}, and quantum logic~\cite{sanaka,gasparoni,
kiesel,walther}. A pulsed source of entangled photons is required
for quantum information protocols for which well-defined timing is
needed to interfere photons from different sources. This becomes
important for experiments involving more than one pair of photons,
such as entanglement swapping~\cite{zukowski,pan}.

There has been much work on designing sources of
polarization-entangled photons that display high visibility
quantum interference~\cite{branning, kim1,kim2, kimpbs, kwiat2,
wong1, shi, nambu, bitton,giorgi, barbieri}. Sources pumped by
continuous-wave (cw) lasers have demonstrated essentially perfect
polarization visibility~\cite{wong1,kwiat2,giorgi}. Sources pumped
by
ultrashort pulses, however, suffer from comparatively low visibility. 
In frequency space, the broader spectrum of a pulsed pump allows
for the generation of down-converted photons whose color provides
distinguishing information about the polarization. This
correlation between polarization and frequency reduces the
polarization visibility~\cite{grice,keller,ou}. Spectral filtering
can be used to eliminate this distinguishing information, while a
thin nonlinear crystal can be used to limit distinguishability
\emph{a priori}. Experiments that rely only on filtering or the
use of thin nonlinear crystals can attain high visibility, but at
lower photon count rates~\cite{giuseppe,grice1,sergienko}. Another
technique to compensate or correct for distinguishable information
is the use of interference~\cite{kwiatprop}. These schemes
interfere distinguishable amplitudes by using two
crystals~\cite{kwiat2, kim1, kim2, bitton}, combining photons at a
beam-splitter ~\cite{kwiatprop, kimpbs, wong1}, or overlapping two
passes through the nonlinear crystal~\cite{branning, shi, giorgi,
barbieri}. We demonstrate a scheme which utilizes the third
concept. The amplitudes to produce a distinguishable (or
non-maximally entangled) photon pair in each pass are interfered
to yield highly entangled states.

Recently, it has been demonstrated that reflecting the pump pulse
and down-converted photon pair back through the nonlinear crystal
stimulates the production of another photon pair in the same mode
as the first~\cite{antia}. This double-pass set-up has generated
four-photon states used to demonstrate spin-1 singlet
states~\cite{howell} and heralded generation of path-entangled
states~\cite{nonlocal}, and additionally to investigate entangled
modes of up to 100 photons~\cite{largen}.
Until now the properties of the two-photon state produced by the
Type-II non-collinear double-pass have not been explored. In this
paper we explain how this double-pass set-up can be used to
optimally generate pulsed entangled photon pairs by effectively
compensating all distinguishing information. Using the double-pass
to interfere distinguishable pair amplitudes we demonstrate
$95.6\%$ visibility  quantum interference 
for 200 femtosecond pulses without the use of narrowband filters,
as well as 98.1\% visibility with 5 nm bandwidth interference
filters. In addition,
this purification scheme allows correlated photons 
emitted in all directions to be collected as
polarization-entangled photons. We conclude by examining the
space-time visibility of the set-up and the effects on multiphoton
states.
\section{VISIBILITY AND COMPENSATION BACKGROUND}
%
The quantum interference visibility we refer to is the
polarization visibility, and it is a common measure of the quality
of a polarization-entangled state. In the frame of the lab the
nonlinear crystal always produces horizontally and vertically
polarized photons, thus the visibility in this basis (HV) is high
regardless of entanglement.
By measuring the correlations in the plus and minus 45 degree
linear (PM) and right and left circular (RL) polarization bases
the quantum entanglement of the state can be determined. The
polarization visibility, as defined here, is
\begin{equation}
\rr{V} = \frac{X_a Y_b + Y_a X_b - X_a X_b - Y_a Y_b}{X_a Y_b +
Y_a X_b + X_a X_b + Y_a Y_b}. \label{eq:vis}
\end{equation}
Here $X_a Y_b$ is the number of coincidence counts measured
between spatial modes $a$ and $b$ having polarizations $X$ and $Y$
respectively, which correspond to HV, PM, or RL depending on the
measurement basis. 
It should be noted that $|\rr{V}| \leq 1$. So, for example, the
$\phi^+$ Bell state has $\rr{V}=-1$ in the HV basis and the
$\psi^-$ Bell state has $\rr{V} = +1$ in all bases.

Before describing our specific experiment we should first explain
what needs to be compensated or filtered to create
maximally-entangled polarization states.  In general we would like
to be able to factorize the polarization from the other degrees of
freedom (such as k-vector and frequency). When that is possible,
gaining information about these observables will not provide any
polarization information. This is not generally the case for
photon pairs emitted from a Type-II source. 
Due to  birefringence there is a temporal delay between the
photons in each mode, and the extraordinary mode walks-off through
the crystal along the optical axis. These temporal and spatial
walk-offs can be effectively eliminated by using compensating
crystals as in Ref.~\cite{kwiatpdc}.

A new problem arises when the crystal is pumped by a broadband
laser: the dispersion and phase-matching conditions in the
nonlinear crystal create differences in the ordinary and
extraordinary spectral modes~\cite{grice, keller,ou}. In the case
of $\beta$ - barium borate (BBO) the spectrum of the ordinary
photons is wider than that of the extraordinary photons. This
disparity increases for increasing pump bandwidth. When using
Type-II non-collinear PDC sources both spatial modes $a$ and $b$
(defined by the intersection of the well-known
rings~\cite{kwiatpdc}) receive both extraordinary and ordinary
photons.  Thus, in principle, the polarization of the photon in
each mode can be determined solely by knowledge of its frequency.
%

\begin{figure}[!t]
\includegraphics[width=\columnwidth]{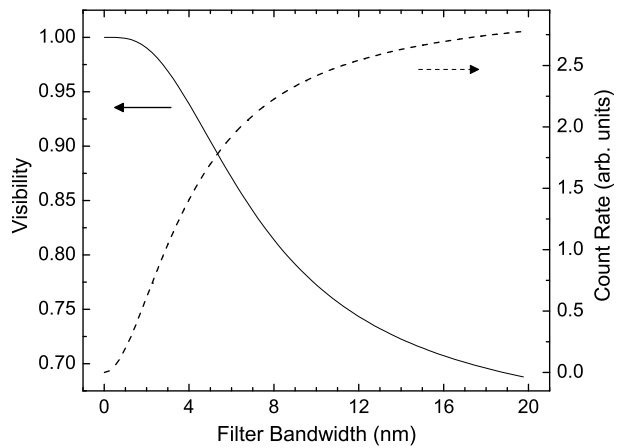}
\caption{Visibility and coincidence count rate as a function of
filter bandwidth. This graph is for a 2 mm thick BBO crystal with
the optical axis at 45 degrees to the pump.  The pump has a
Gaussian profile with full width half-max of 1 nm around 390 nm,
and the down-converted photons are degenerate around 780 nm.  The
filter bandwidth is the half-width at $1/e$ max.}
\label{fig:filtergraph}
\end{figure}

In the following we concentrate on spectral mode
dististinguishablity. As mentioned above, the spectra of the
down-converted photons can be made indistinguishable by filtering
tightly, however this causes the photon count rates to decrease as
well. Following a calculation by Grice and Walmsley~\cite{grice},
the polarization visibility and coincidence count rate can be
obtained as functions of the filter bandwidth. Figure
~\ref{fig:filtergraph} shows these functions for typical crystal
and pump laser parameters in our experiments. The visibility
without filters is 62.2\%, but this can be improved to 90.8\% by
using 5 nm bandwidth filters centered at 780 nm. If one would like
to obtain 95\% visibility, then the count rate will be half of the
unfiltered rate (assuming filters with ideal peak transmission).

Another technique to eliminate the spectral mismatch is the use of
a thin nonlinear crystal to increase phase-matching (e.g., length
$<$ 0.5 mm for our experimental parameters).  This results in both
extraordinary and ordinary photons having identical, broad spectra
(FWHM $>$ 15 nm). The use of a thin crystal (100 $\mu$m) has been
demonstrated to give high visibility fringes ($\rr{V} \simeq $
98\%), however the resulting coincidence count rates were below
100 Hz~\cite{sergienko}. More recent experiments have attained
much higher count rates using a thin crystal pumped by a cw
laser~\cite{lee}.
%
%

A number of experiments have utilized interference in order to
compensate PDC~\cite{branning, kim1, kim2, kimpbs, kwiat2, wong1,
shi, nambu, bitton,giorgi, barbieri}. The first proposal for such
a technique was by Kwiat, \emph{et al.} who suggested using a
half-waveplate and polarizing beam-splitter (PBS) to combine the
output of two Type-II collinearly phase-matched
down-converters~\cite{kwiatprop}.
This technique, combining photons at a PBS, is the one of three
general types of interference compensation schemes. The others are
overlapping the emission amplitudes of two nonlinear crystals and
reflecting the down-converted photons and pump back through the
nonlinear crystal. The first two techniques have proven quite
effective by demonstrating visibility between 90-92\% for
ultrashort pulses without the use of spectral
filters~\cite{kimpbs,kim1,kim2,nambu}. The pulsed, double-pass
schemes reported~\cite{branning,shi} also increase visibility to
64\% and 80\%, respectively, without filtering. Specific
comparisons become difficult due to the various types of
measurements performed and the different experimental conditions
(e.g., phase-matching, pump pulse length, crystal length). The
double-pass scheme presented here provides high visibility for
unfiltered pulses without the need to post-select, while using the
same set-up that has reliably produced multiphoton entangled
states~\cite{howell, antia, nonlocal, largen}. In the following
section we discuss the theory behind the double-pass compensation
scheme employed in our experiment.
\section{Theoretical Model}
%
To explain how the double-pass scheme interferes distinguishable
amplitudes to produce a higher visibility state we present a
simplified version of the detailed theory for PDC. Typically one
begins with the effective interaction Hamiltonian for Type-II
non-collinear single-pass PDC
\begin{eqnarray}
 \rr{H}_{\rr{Int}}(t) \!\!\!\!  &=&
\nonumber \\
& i \hbar \kappa &\!\!\!\! \int \!\!\! \int \rr{d}\omega_{o} \,
\rr{d}\omega_{e} \big (\hat{a}_{h}^{\dag}(\omega_o)
\hat{b}_{v}^{\dag}(\omega_e) \, - \, \hat{a}_{v}^{\dag}(\omega_e)
\hat{b}_{h}^{\dag}(\omega_o) \big)
\nonumber \\
& \times & \!\! f(\omega_o, \omega_e)+ h.c. \label{eq:hint}
\end{eqnarray}
%
where $f(\omega_o,\omega_e)$ is the joint spectral amplitude
function~\cite{grice, keller,ou, branning}. The interaction
parameter, $\kappa$, depends on the pump intensity, the crystal's
$\chi^{(2)}$ nonlinearity, and the crystal length.

This detailed approach can be somewhat cumbersome, and a simpler
technique is to ignore frequency correlations between photons.
The frequency correlations do not affect the two-photon
visibility, however they do become important when using more than
one photon pair~\cite{grice2}.  Ignoring frequency correlations
means that $f(\omega_o,\omega_e) = g(\omega_o) h(\omega_e)$, and
the frequency integrals above can be evaluated separately
resulting in
\begin{equation}
\rr{H_{1 pass}} = i \hbar \kappa (\hat{a}_{h \,}^{\prime \,\dag}
\hat{b}_{v \,}^{\dag} \, - \, \hat{a}_{v \,}^{\dag} \hat{b}_{h
\,}^{\prime \,\dag})+ h.c.
\end{equation}
The prime indicates that the ordinary photons are emitted into a
different spectral mode than the extraordinary photons. The prime
mode can be decomposed as $\hat{a}^{\prime}= x \hat{a}_{1} +
\sqrt{1-|x|^2}\hat{a}_{2}$ where modes 1 and 2 are orthogonal
($[\hat{a}_{1},\hat{a}_{2}^{\dag}]=0$). This characterizes the
partial overlap between the ordinary and extraordinary spectral
modes parameterized by the value of $x$ (the only parameter of
this simplified model). For the remainder of the paper we take
``ordinary'' and ``extraordinary'' to label \emph{spectral modes}
rather than polarization modes. The Hamiltonian can then be
written
\begin{eqnarray}
\rr{H_{1 pass}} \!\!&=& \! i \hbar \kappa \big[x
\bellst{h1}{v1}{v1}{h1}
\nonumber \\
&+& \!\! \sqrt{1-|x|^{2}} \bellst{h2}{v1}{v1}{h2}\big]+ h.c.
\label{eq:1pass}
\end{eqnarray}
The visibility of the photon pairs generated from such a
Hamiltonian will be perfect (V=1) in the HV basis. In the PM and
RL polarization bases, however, the visibility will be limited to
$x^{2}$. The second term does not contribute to the visibility
because the terms $a_{h2} b_{v1}$ and $a_{v1} b_{h2}$ will not
interfere when measured in the PM or RL basis. Although the value
of $x$ can only be predicted by the more detailed theory,
Eq.~(\ref{eq:1pass}) conveniently explains the features of the
double-pass as described below.

\begin{figure}[!t]
\includegraphics[width=\columnwidth]{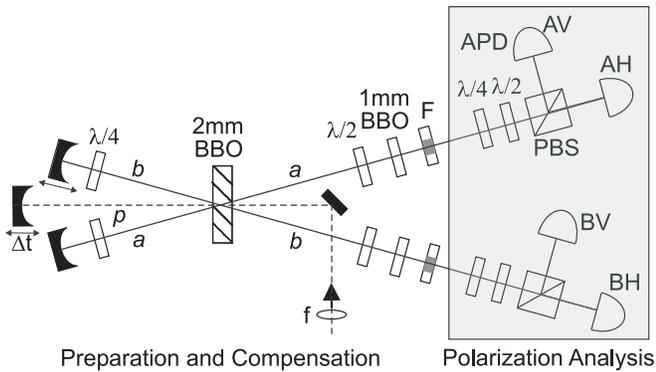}
\caption{Experimental set-up.  The pump beam (dotted line) is
focused by a lens (f, focal length = 50 cm), and passes through
the nonlinear crystal (BBO) twice with a variable delay. The
down-converted photons (solid line) from the first pass have their
polarization rotated by 90 degrees by twice traversing $\lambda/4$
waveplates at 45 degrees. After the second pass the photons are
compensated for walk-off (see Fig.~\ref{fig:walkoff}) and filtered
(F) before entering single-mode fibers for polarization analysis.}
\label{fig:setup}
\end{figure}

The effect of the double-pass is to add another set of modes, 3
and 4. These are analogs to 1 and 2, distinguished only by their
time of arrival at the detectors. The polarization of the
down-converted photons from the first pass is flipped
(H$\rightleftarrows$V) before they are reflected back into the
crystal (Fig.~\ref{fig:setup}). The resulting Hamiltonian reads
\begin{eqnarray}
 \rr{H_{2 pass}} \!\! &= \, i \hbar \kappa
\Big[& \!\! x  \bellst{v1}{h1}{h1}{v1}
\nonumber \\
&+& \!\!\!\!\!\!\!\! \sqrt{1-|x|^{2}} \bellst{v2}{h1}{h1}{v2}
\nonumber \\
&+& \!\!\!\!\!\!\!\!  e^{i \omega \Delta t} \Big( x
\bellst{h3}{v3}{v3}{h3}
\nonumber \\
&+& \!\!\!\!\!\!\!\!  \sqrt{1-|x|^{2}} \bellst{h4}{v3}{v3}{h4}
\Big) \Big]+ h.c., \label{eq:2passb}
\end{eqnarray}
where $\omega$ is the frequency of the pump laser, and $\Delta t$
is the difference in arrival times of the two passes at the
detectors . The change in polarization labels on the first pass
can alternatively be viewed as an exchange of the spectral-mode
labels ($1 \!\! \rightleftarrows \! 2$). When $\Delta t$ is much
less than the coherence time of the down-converted photons, modes
1 and 3 are equivalent, and modes 2 and 4 are equivalent. The
Hamiltonian can then be written
\begin{eqnarray}
 \rr{H_{2 pass}^{\Delta t \simeq 0}} \!\! &= i \hbar \kappa
\Big[&x (1+e^{i \theta})\bellst{h1}{v1}{v1}{h1}
\nonumber \\
&+& \!\!\!\!\!\!\!\! \sqrt{1-|x|^{2}}\Big( (e^{i
\theta}\hat{a}_{h2}^{\dag}\hat{b}_{v1}^{\dag} \, - \,
\hat{a}_{v2}^{\dag} \hat{b}_{h1}^{\dag})
\nonumber \\
&+& \!\!\!\!\!\!\!\! (\hat{a}_{h1}^{\dag}\hat{b}_{v2}^{\dag} \, -
\, e^{i \theta} \hat{a}_{v1}^{\dag} \hat{b}_{h2}^{\dag})
 \Big) \Big]+ h.c.,
\label{eq:2passbzero}
\end{eqnarray}
where $\theta = \omega \Delta t$ is the relative phase of the two
passes. Rotating the polarization of the first pass and
overlapping it with the second pass creates a coherent
superposition of down-converting in each pass. The first pass
provides ordinary V-polarized photons and extraordinary H photons,
while the second pass creates ordinary H and extraordinary V
photons. When both passes are combined, knowing the spectral mode
no longer provides any polarization information. This is
equivalent to symmetrizing $f(\omega_o,\omega_e)$ with respect to
interchanging its arguments~\cite{branning}. Through this
technique we optimally compensate the spectral mismatch of pulsed
PDC.

Notice that the first term of Eq.~(\ref{eq:2passbzero}) can be
stimulated (or suppressed) as a function of the phase ($\theta$).
The other terms have a constant amplitude since the phase appears
inside these terms. When the phase is equal to zero, the state
produced by Eq.~(\ref{eq:2passbzero}) is a superposition of three
$\psi^-$ Bell states, each with visibility equal to one.

Even from this simple, two-mode model it becomes evident that this
technique only works perfectly on two-photon, PDC
events~\cite{note1}. Multiphoton events ($n>2$) result from the
higher-order terms of the interaction Hamiltonian. Exponentiating
the Hamiltonian mixes the modes such that the spatial and spectral
mode labels again become distinct. For the simplest example,
consider the second-order (four-photon) state. Once
Eq.~(\ref{eq:2passbzero}) is squared only the terms that square
with themselves remain maximally entangled; the cross terms are
multimode in the spectral mode space. Nonetheless, the double-pass
scheme does mitigate the spectral mismatch, and multiphoton
experiments using this set-up have reported four-photon
visibilities over 80\% ~\cite{nonlocal}. The primary source of
decoherence in those experiments was distinguishable temporal
modes imparted by the short down-converted pulse length compared
to that of the pump pulse~\cite{grice2, ou2}.
\section{Experimental Demonstration}
\subsection{Double-Pass Set-Up}
\begin{figure}[!t]
\includegraphics[width=\columnwidth]{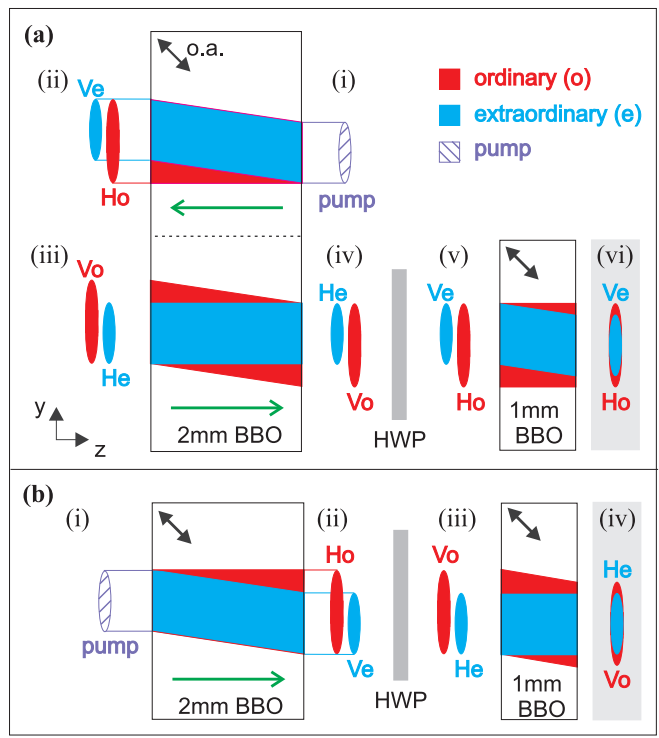}
\caption{(Color online) Walk-off compensation. Long arrows
indicate direction of beam propagation. The shaded boxes depict
the final states for each pass. (a) Compensation of photons
created in the first pass. (i) The vertically-polarized pump beam
walks off along the optical axis (o.a.) of the 2 mm crystal
creating down-converted photons along its entire length. The
V-polarized, extraordinary ($e$) photons walk with the pump beam,
while the H-polarized, ordinary ($o$) photons continue along the
z-axis. (ii) The resulting $o$ spatial profile is stretched along
the y-axis (perpendicular to optical table) and lags behind in
space-time (z-axis) by 1 mm of birefringence on average. (iii) The
net effect of twice traversing the quarter-waveplates and
reflecting from the curved mirrors (not shown) is to flip the
polarizations and invert the spatial profiles. (iv) The second
trip through the 2 mm crystal (shifted for clarity) walks the
V-polarized, $o$ photons along the o.a. They also emerge ahead in
time. (v) A HWP flips the polarizations. (vi) A 1 mm BBO crystal
walks the $o$ and $e$ photons together in space and time. (b)
Second-pass photon compensation. (i) The pump pulse passes through
the 2 mm BBO a second time (ii) generating down-converted photons.
(iii) A HWP flips polarizations. (iv) The 1 mm BBO again overlaps
the $o$ and $e$ photons.  Notice that the final state of the first
pass photons corresponds to that of the second pass, however the
polarizations are interchanged.}
 \label{fig:walkoff}
\end{figure}
To test the double-pass compensation scheme we utilized a slightly
modified version of the design of Ref. ~\cite{antia}.  A
frequency-doubled, Ti:Sapphire laser operating at 82 MHz
repetition rate provides 200 fs pulses centered at 390 nm. As
illustrated in Fig.~\ref{fig:setup} these pulses are focused onto
a 2 mm thick BBO crystal where PDC takes place. The pump pulse and
down-converted photons leave the crystal in distinct spatial modes
$p$, $a$, and $b$.  Each mode is aligned normal to the center of a
curved mirror, which reflects the modes back into the crystal
along their same paths. Modes $p$ and $b$ are each on translation
stages so that the arrival time of all three modes can be matched
at the crystal. Quarter-waveplates oriented at 45 degrees are
placed into modes $a$ and $b$. These flip the polarization of the
first-pass photon amplitudes when traversed twice. When the timing
of $p$, $a$, and $b$ match, these amplitudes interfere with those
of creating a photon pair in the second pass to create the
two-photon, high-visibility state of Eq.~(\ref{eq:2passbzero}).

After the second pass through the crystal, modes $a$ and $b$ pass
through half-wave plates at 45 degrees and 1 mm BBO crystals to
correct for spatial and temporal walk-off. The entire walk-off
compensation scheme is shown schematically in
Fig.~\ref{fig:walkoff}, and is described in detail in the figure
caption.  The result is that photons from the first pass with
swapped mode labels are completely indistinguishable from those
produced in the second pass with normal mode labels. 
The temporal profiles overlap and the spatial profiles are
centered to facilitate equal fiber coupling. Once the photons are
properly compensated, depending on the measurement being
performed, they are filtered spectrally with narrowband or
edgepass filters and spatially with single-mode fibers. The
visibility is measured using the polarization analysis set-up
consisting of a quarter-waveplate, half-waveplate, and polarizing
beam-splitter (PBS).  The light is detected by avalanche
photodiodes (APDs) connected to a coincidence logic circuit with a
window of 2 ns.

Mode distinguishability is not the only source of decreased
visibility. More than one PDC event may take place per pump pulse
with an amplitude proportional to the interaction parameter
$\kappa^{n}$ for $n$ pair events. These events can degrade the
two-photon visibility~\cite{largen}. This, however, has a smaller
effect on the visibility than mode distinguishability at the pump
powers used in our experiment. For small values of $\kappa$ the
contribution from four-photon events is the dominant noise term,
and the visibility decreases linearly. 
The visibility data presented in this paper depicts this trend
showing a typical 4-8\% visibility decrease as the pump power is
increased by a factor of order 100.
\begin{figure}[!t]
\includegraphics[width=\columnwidth]{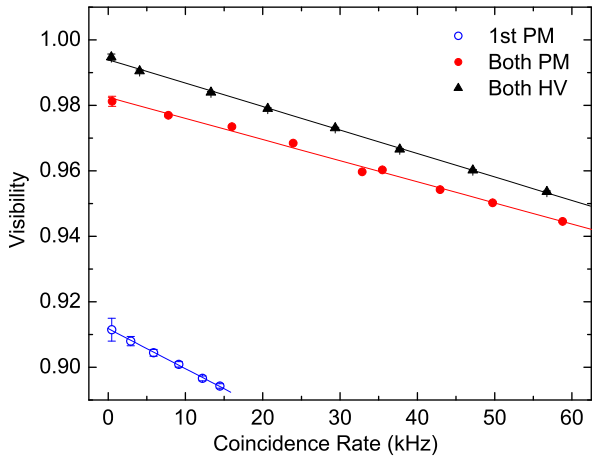}
\caption{(Color online) Visibility as a function of coincidence
count rate with 5 nm bandwidth filters.  The visibility decreases
linearly as the pump power is increased in both HV (triangles) and
PM (circles) bases due to increasing noise from four-photon,
down-conversion events. Record high visibilities for pulsed
down-conversion of up to 98.1\% are shown in the PM basis using
the double-pass; an improvement of 7.0\% over the 1st pass alone.
The highest count rates correspond to a pump power of 400 mW.}
\label{fig:intfilters}
\end{figure}
\subsection{Spectral Compensation Results}
The spectral compensation was tested experimentally by comparing
the visibility of the individual passes to that of the double-pass
with and without narrowband filters.  The single-pass visibility
is given by $x^2$, which can be determined by
calculation~\cite{grice}, or by directly measuring the visibility
of a single pass. The $x^2$ values listed below are determined by
measurement, but they are in good agreement with theory (cf.
Fig.~\ref{fig:filtergraph}). The double-pass visibility predicted
by Eq.~(\ref{eq:2passbzero}) should be equal to one at phase equal
to zero, regardless of $x$.
\begin{figure}[!t]
\includegraphics[width=\columnwidth]{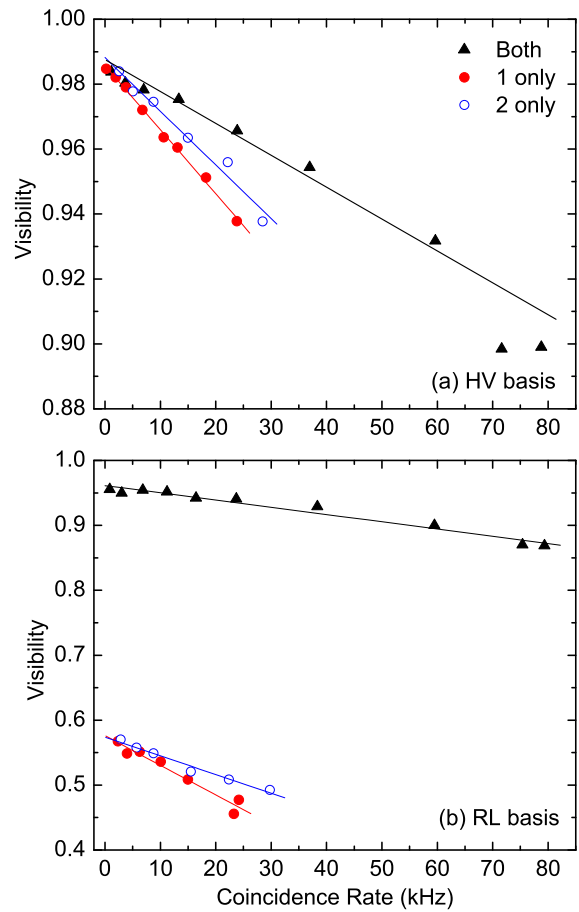}
\caption{(Color online) Visibility as a function of coincidence
count rate with only edgepass filtering.  The data displayed is
for each pass individually (circles) and both passes combined
(triangles). Lines are provided to guide the eye. (a) HV basis.
The visibility is high in all three cases, and decreases at higher
power due to noise from four-photon events. (b) RL basis. The
double-pass vastly improves the visibility from around 55\% for
single passes to over 95\%. Error bars are smaller than the height
of the symbols. The highest count rates correspond to a pump power
of 300 mW.} \label{fig:coffilters}
\end{figure}

When using narrowband filters the spectral modes are governed by
the convolution of the ordinary and extraordinary spectral
profiles with those of the filters, and are therefore almost
indistinguishable, ($x^2=.91$). The visibility of the double-pass
with 5 nm filters as a function of the coincidence count rate is
shown in Fig.~\ref{fig:intfilters}. The count rate was decreased
by rotating the pump polarization in steps, which reduced the pump
power along the optical axis of the crystal. A maximum visibility
of $98.1\pm0.15\%$ in the PM basis is considerably higher than the
91\% predicted and measured for a single-pass. The count rate at
this point is 519 Hz.
The double-pass also allows high count rates and visibility
simultaneously by displaying over 50000 coincidences per second
with 95\% visibility. This corresponds to over five million
entangled photon pairs per second produced by the crystal when our
10\% collection efficiency is considered.

When only edgepass filters are used to eliminate residual UV light
from the pump beam, the ordinary and extraordinary spectra differ
and the value of $x$ is small ($x^{2} \approx .57$).  This can be
seen in the RL visibility of the individual passes in
Fig.~\ref{fig:coffilters}(b). The individual pass visibilities
peak at 57\%, however by interfering these same two passes the
visibility jumps as high as $95.6\pm0.45\%$ at a rate of 855 Hz.
The double-pass has effectively eliminated the spectral
distinguishability without having to sacrifice photons to lossy
narrowband filters.
\subsection{Spatial Compensation}
%
To this point we have focused on the double-pass compensating the
spectral mismatch, 
however this scheme works comparably well in eliminating
spatial-mode mismatch.  To verify this we aligned the collection
optics such that modes $a$ and $b$ corresponded to conjugate
points slightly away from the intersection of the rings. This
varies the amplitude of the individual terms ($a_h b_v$ and $a_v
b_h$) in each pass, parameterized by $y$ below
\begin{eqnarray}
\rr{H_{misalign}} \!\!\! &=& \!\! i \hbar \kappa \big [(y
\hat{a}_{h1}^{\dag} \hat{b}_{v1}^{\dag} \, - \,  \sqrt{1-y^2}
\hat{a}_{v1}^{\dag} \hat{b}_{h1}^{\dag})
\nonumber \\
&+& \!\! e^{i \omega t} (y \hat{a}_{v2}^{\dag} \hat{b}_{h2}^{\dag}
\, - \, \sqrt{1-y^2} \hat{a}_{h2}^{\dag} \hat{b}_{v2}^{\dag})\big
] + h.c. \label{eq:misalign}
\end{eqnarray}
Mode 1 labels the first pass through the crystal, and 2 the
second, where the same misalignment, $y$, is assumed for both.
Here we have ignored the spectral differences parameterized by
$x$, but it is trivial to show that compensation holds given both
spatial
and spectral mismatch. 

If the passes overlap in time, then they can interfere and produce
an entangled state upon leaving the crystal the second time
\begin{eqnarray}
\rr{H_{misalign}^{\Delta t\simeq0}} \!\!\! &=& \!\! i \hbar \kappa
\big [y \bellstpb{h}{v}{v}{h}
\nonumber \\
&-& \!\! \, \sqrt{1-y^2} \bellstpa{h}{v}{v}{h} \big ] + h.c.
\label{eq:misalignzero}
\end{eqnarray}
Once again the two passes have been interfered to produce a Bell
state with $|\rr{V}|=1$, regardless of the value of $y$. This
means that even if the crystal only emits $a_h b_v$ ($y=1$) in a
single pass, maximal entanglement can still be obtained with the
double-pass. All conjugate points of the Type-II rings can be
collected as sources of polarization-entangled photons, rather
than just the intersection of the rings. This allows for an
increase in photon flux while preserving high visibility (for
instance, with the use of multimode collection).

To verify this experimentally, the count rates of the individual
passes were aligned to two different $y$ values: $y^2 = .28$ and
$y^2 = .12$. The average single-pass visibilities at these points
were 47.9\% and 23.5\%,
respectively, with 5 nm interference filters. 
When both passes were combined, the measured visibility jumped to
94.2\% 
and 91.8\% 
, respectively, in the PM basis. Similar results were obtained in
the RL basis. Coincidence count rates were over 5000 Hz for both
measurements. These high visibilities indicate that substantial
polarization entanglement exists for any photons with correlated
k-vectors emitted from the crystal.

\section{Space-Time Visibility}
%
So far we have restricted our definition of visibility to
polarization visibility, obtained by comparing the signal of terms
consistent with the state desired to noise from unwanted terms.
Another type of measurement that can be done with the double-pass
is the space-time, or path visibility. This is the visibility of
the fringes obtained when the pump mirror is translated along the
axis of the pump beam.  This varies the phase between the two
passes, that is $\theta$ of Eq.~(\ref{eq:2passbzero}), resulting
in sinusoidal oscillations in the down-converted, photon-pair
intensity. 

Using the simple model described in Section III predictions can
also be made for the path visibility for given values of $x$ and
$y$. We consider a well-aligned set-up with imperfect filtering
($y=1$, $x<1$). Interestingly, when the polarization is analyzed
in the HV basis the path visibility is limited to $x^2$, but when
analyzed in the PM or RL bases the visibility is one, regardless
of $x$. This feature of the system is elucidated by the
simplified interaction Hamilitonian.

To see these properties consider Eq.~(\ref{eq:2passbzero}) for
$x<1$. Of the three terms, only the first has a phase-dependent
amplitude. This term oscillates as a function of phase to produce
interference fringes.  The other terms are phase independent, and
therefore degrade the path visibility in the HV basis by providing
a constant background of HV counts. It is easy to show that
evaluating the HV fringe visibility with this constant background
gives a maximum value of $x^2$. Upon rotating to the PM (or RL)
basis all of the terms that contribute to the PM (RL) coincidences
have a phase-dependent amplitude: ($1 + e^{i \theta}$). Therefore,
the PM (RL) count rate is zero at $\theta = n \pi$ (for odd
integer $n$) regardless of $x$, and the PM (RL) space-time
visibility is one.

\begin{figure}[!t]
\includegraphics[width=\columnwidth]{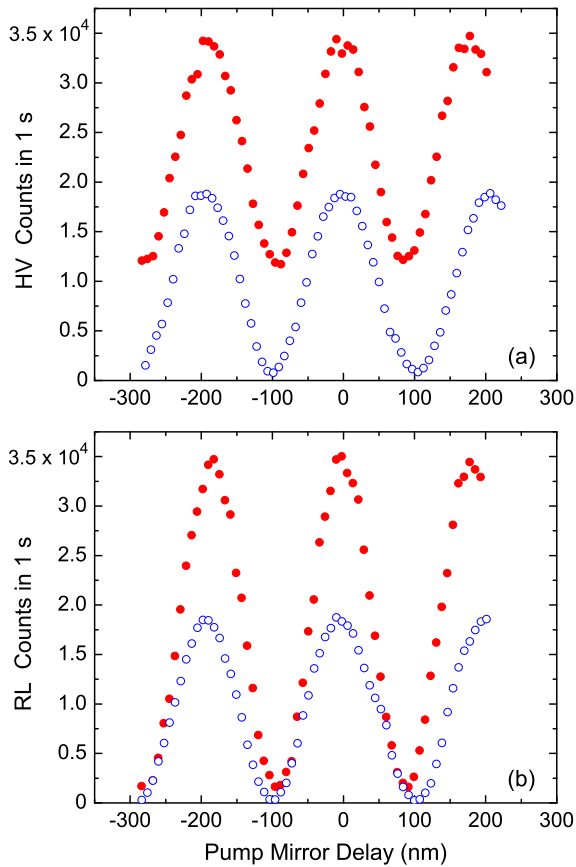}
\caption{(Color online) Space-time visibility for spatially
aligned passes.  The visibility is measured with 5 nm bandwidth
filters (open circles) and edgepass filters (closed circles). In
the (a) HV basis visibility is limited to $x^2$, however in the
(b) RL basis the visibilities are close to the theoretical value
of 1. The fringe visibility increases the closer the minima get to
zero counts.} \label{fig:fringes}
\end{figure}

These features are illustrated in Fig.~\ref{fig:fringes}.
Figure~\ref{fig:fringes}(a) shows the interference fringes of HV
coincidences measured with narrowband filters ($x^2 = .91$) and
with edgepass filters ($x^2 \approx .49$). The visibility is
determined by a fit of the data with a sine-squared function. The
HV fringe visibility is 91.3\% with 5 nm filters compared to
48.2\% with only edgepass.  In contrast, the RL inteference
fringes (Fig.~\ref{fig:fringes}(b)) display 96.1\% visibility with
5 nm filters and 92.4\% with edgepass filters.  Similar results
were obtained in the PM basis.

As predicted the visibility in the RL basis remains high
regardless of the spectral-mode distinguishability (value of $x$),
while in HV it goes as $x^2$. These fringe visibilities are
affected adversely by multiple-pair events in a similar manner as
the polarization visibility.  This limits the maximum visibilities
in all bases as a function of the interaction parameter and
detection efficiencies.
\section{Summary and Discussion}
While the double-pass scheme has many advantages, a serious
challenge is the required interferometric stability. Unlike
four-photon experiments with the
double-pass~\cite{antia,howell,nonlocal}, the phase does not just
modulate the intensity of the state, but also the phase within
some terms (see Eq.~(\ref{eq:2passbzero})). This means that not
only does the signal decrease away from $\theta = 0$, but the
state produced contains components that are not $\psi^-$ states.
Typical stability periods before the interferometer would drift
approximately 3\% from the maximum, were around 2 minutes in an
environment partially isolated from external vibrations.  All
polarization visibility data was collected at spatial-fringe
maxima for intervals shorter than this period. A clever technique
to increase stability is to replace the three separate curved
mirrors with a single, large curved mirror with high reflectivity
at both the pump and down-converted photon wavelengths. This has
been demonstrated for a Type-I non-collinear geometry in
Ref.~\cite{giorgi, barbieri}. This single-arm interferometer then
makes it possible to build a reference interferometer as in
Ref.~\cite{branning} for active stabilization.

We have presented the theory and experimental results for a
double-pass compensation scheme for pulsed parametric
down-conversion. A simplified Hamiltonian was introduced to
explain the form of the down-converted state with and without
compensation. When the passes overlap the amplitudes for normally
labelled photons and those with switched labels become
indistinguishable, thereby optimally compensating any spectral and
spatial-mode mismatch.

Using the double-pass scheme high polarization visibility was
achieved without spectral filters (95.6\%), and even higher
(98.1\%) with 5 nm bandwidth filters.  The lower visibility
without filters may have resulted from the reduced effectiveness
of the optics with a broader spectrum of photons. The waveplates
and compensating crystals no longer put the appropriate phases on
photons as their wavelengths differ from 780 nm. Another
possibility is the effect of group velocity dispersion, which was
not considered in our model. Double-pass compensation is not
limited to spectral-mode distinguishability. Spatial-mode mismatch
was eliminated by combining single passes with an average
visibility of 23.5\% to attain over 92\% visibility.

In addition to these demonstrations, we verified a prediction
about the space-time visibility based on the simplified theory. In
contrast to polarization visibility, the space-time visibility
measured in the HV basis is reduced by spectral mismatch, while
the visibility remains one for the complementary PM and RL bases.
The space-time and spatial-mode mismatch visibilities were limited
primarily by two-pair, down-conversion events, since this data was
taken only at high pump power.

This research has been supported by NSF Grant No. 0304678 and by
the DARPA MDA 972-01-1-0027 grant. J.F.H. thanks Lucent
Technologies CRFP for financial support. We acknowledge support
from Perkin-Elmer regarding the SPCM-AQR-13-FC single-photon
counting modules.


\begin{thebibliography}{99}
\bibitem{note1}
The non-ideal compensation is evident from the simplified model,
although the assumption in deriving that model (about being
justified in ignoring frequency correlations) is invalid. The
joint spectral amplitude function can no longer be assumed to be
separable~\cite{grice2}.
\end{thebibliography}
\end{document}